\documentclass[twocolumn,aps,pre,superscriptaddress,floatfix]{revtex4-1}
\usepackage{graphicx}
\usepackage{epsfig}
\usepackage{thumbpdf}
\usepackage{amsfonts}
\usepackage{times}     
\usepackage{float}
\usepackage{indentfirst}   %
\usepackage{amsmath}
\usepackage{epstopdf}
\usepackage{textcomp}
\usepackage{tocvsec2}
\usepackage{epsfig}
\usepackage{xcolor}
\usepackage{comment}
\usepackage{CJK} 

\usepackage[section]{placeins}
\makeatletter\AtBeginDocument{%
     \expandafter\renewcommand\expandafter\subsection\expandafter
       {\expandafter\@fb@subsecFB\subsection}%
     \newcommand\@fb@subsecFB{\FloatBarrier
     \gdef\@fb@afterHHook{\@fb@topbarrier \gdef\@fb@afterHHook{}}}
     \g@addto@macro\@afterheading{\@fb@afterHHook}
     \gdef\@fb@afterHHook{}
  }
\usepackage[colorlinks=true,linkcolor=black]{hyperref} 

\newcommand{\av}[1]{\left\langle {#1} \right\rangle}
\newcommand{\be}{\begin{equation}}
\newcommand{\ee}{\end{equation}}

\begin{document}
\begin{CJK}{UTF8}{gbsn}

\title{Mutually cooperative epidemics on power-law networks}

\author{Peng-Bi Cui (崔鹏碧)} 
\affiliation{Istituto dei Sistemi Complessi (ISC-CNR), UOS Sapienza, Piazzale A. Moro 2, 00185 Roma, Italy}
\affiliation{Web Sciences Center, University of Electronic Science and Technology of China, Chengdu 611731, China}
\affiliation{Big Data Research Center, University of Electronic Science and Technology of China, Chengdu 611731, China}

\author{Francesca Colaiori}
\affiliation{Istituto dei Sistemi Complessi (ISC-CNR), UOS Sapienza, Piazzale A. Moro 2, 00185 Roma, Italy}
\affiliation{Dipartimento di Fisica, Sapienza Universit\`a di Roma, Roma, Italy}

\author{Claudio Castellano}
\affiliation{Istituto dei Sistemi Complessi (ISC-CNR), Via dei Taurini 19, 00185
 Roma, Italy}

%

\begin{abstract}
The spread of an infectious disease can, in some cases, promote
the propagation of other pathogens favouring violent outbreaks, 
which cause a discontinuous transition to an endemic state. 
The topology of the contact network plays a crucial role in
these cooperative dynamics. 
We consider a susceptible--infected--removed (SIR) type model with two 
mutually cooperative pathogens: an individual already infected with one
disease has an increased probability of getting infected by the other. 
We present an heterogeneous mean-field theoretical approach to 
the co--infection dynamics on generic uncorrelated power-law degree-distributed 
networks and validate its results by means of numerical simulations. 
We show that, when the second moment of the degree distribution is finite,
the epidemic transition is continuous for low cooperativity, 
while it is discontinuous when cooperativity is sufficiently high.
For scale-free networks, i.e. topologies with diverging second moment,
the transition is instead always continuous.
In this way we clarify the effect of heterogeneity and
system size on the nature of the transition and we validate the physical 
interpretation about the origin of the discontinuity.
\end{abstract}

\maketitle
\end{CJK}

\section{Introduction}
\label{sec:introduction}
Modelling epidemic dynamics plays a key role in predicting disease
outbreaks and designing effective strategies to prevent or control
them.  While traditional theories of disease propagation ignore
network effects, there has been a large amount of research in the
past decades aimed at understanding how the topological structure of
contact networks affects the dynamics upon them~\cite{PastorSatorras2015}.
Such studies have focused mainly on the dynamics of a single epidemic 
disease.

An issue of growing interest in current epidemiological research is
how concurrent spreading diseases interact with each other.
Such coupled spreading scenarios can occur when either multiple 
pathogens or multiple strains of the same disease simultaneously 
propagate in the same population.
The complexity of the problem is largely increased
already in the two--pathogen case.  Two diseases circulating in the
same host population can interact in many different ways, with either
synergistic or antagonistic effects.

A well known type of interaction is cross--immunity: an
individual infected with one disease becomes partially or
fully immune to infection by the second one. 
In this case the two pathogens compete for the same population 
of hosts.  The competition between epidemics that are mutually 
exclusive or antagonistic was studied
in~\cite{Newman2005,Funk2010,Marceau2011,Miller2013}.

An opposite case that is recently gaining attention
is the spreading of two or more cooperating pathogens: in
this case an individual that is already infected with one disease
has increased chance of getting infected with another.  A notable example
is the 1918 Spanish flu pandemic that involved about one--third of the
world's population killing tens of millions of people within months
~\cite{Taubenberger2006}.
A considerable proportion of the infected were co--infected by
pneumonia~\cite{Morens2008,Brundage2008}, and most deaths where caused
not directly by the virus, but by the secondary bacterial infection.
Another well--known case is that of HIV, which increases the host
susceptibility to other pathogens, in particular to the hepatitis C
virus (HCV)~\cite{Sulkowski2008}.

In co--infections, positive feedback between multiple diseases can
lead to more rapid outbreaks.  One of the most interesting questions
is whether cooperation can change the epidemic
transition from being continuous to abrupt when external conditions
vary, even slightly, as for a microscopic change in infectivity.  This
is an extremely relevant problem since the possibility to enact
countermeasures critically depends on the type of transition: in the
discontinuous case the epidemic can take over a population explosively
without any of the warning signs which occur in single epidemics.

Recently~\cite{NewmanFerrario2013} Newman and Ferrario introduced a
model of co-infection based on the susceptible--infected--removed (SIR)
model~\cite{Kermac1927} with two diseases diffusing over the same
contact network. One disease spreads freely while the second can only
infect individuals already infected with the first.  The authors
assume total asymmetry and time--scale separation in the spreading of
the two pathogens.  The system displays two epidemic thresholds, the
first one being the usual SIR threshold, the second occurring
when the fraction of the population infected with the first disease
becomes large enough to allow the spread of the second.  No
qualitative changes in the nature of the outbreaks appear.
In~\cite{Chen2013} Chen, Ghanbarnejad, Cai, and Grassberger (CGCG)
introduced a generalized SIR model (CGCG model) to include {\it
mutual} cooperative effects of co--infections.  In the CGCG model
two different diseases simultaneously spread in a population. Having
being infected with one disease gives an increased probability to get
infected by the other.
The amount of this increase is a proxy of the mutual {\it
cooperativity} between the two diseases.  The authors study the
model at mean--field level and observe that cooperative effects,
depending of their strength, can cause a change of the transition 
from continuous to discontinuous. 
In~\cite{Janssen2016} Janssen and Stenull showed that the CGCG model
is equivalent, in mean--field, to the homogeneous limit of an
extended general epidemic process and clarify the spinodal nature 
of the discontinuous transition observed.  
In~\cite{Chen2015,Chen2016} CGCG analyzed how in their model 
the type of transition depends on the contact network topology by
simulating the CGCG model on both random networks and lattices.
They concluded that a necessary condition for discontinuous 
transitions to occur, when starting from a doubly-infected node, 
is the relative paucity of short loops with respect to long ones. 
More in detail they argue that a discontinuous transition
occurs if the two epidemics first evolve separately and, only
after the independent clusters of singly-infected nodes
have reached endemic proportions, the two epidemics meet: at that
point cooperativity implies that both clusters rapidly become 
doubly infected. A necessary condition then is that there are
few short loops (otherwise the two pathogens immediately cooperate
and the transition is continuous) and
there are long loops (otherwise cooperativity has no effect and
one sees only single infections). In agreement with this scenario
CGCG do not observe discontinuous transitions on trees and
on 2--d lattices, while they do observe them on Erd{\"o}s--R{\'e}nyi (ER)
networks, on 4--d lattices, and on 2--d lattices with sufficiently
long-range contacts.  On 3--d lattices the existence of discontinuous
transitions depends on the details of the microscopic realization of 
the model.
Moreover, the observed discontinuous transitions are of hybrid
type~\cite{Goltsev2006,Parisi2008}, i.e., exhibiting also some features
of continuous ones.
For other recent work about cooperating infections see 
Refs.~\cite{Sanz2014,AzimiTafreshi2016,Brockmann2016}.

An open question left by the analysis in~\cite{Chen2015,Chen2016}
has to do with what happens for co--infections on generic broadly
degree-distributed networks.
Simulations performed on a Barabasi--Albert (BA) topology seem
to indicate a continuous transition even for strong cooperativity.
However, it is not clear whether these results are affected
by finite size effects and what happens for co--infections on other broadly 
distributed networks. Theoretical 
approaches~\cite{Chen2013, Janssen2016} deal only with homogeneous
networks.

In this paper we elucidate these issues.
We first numerically show that for power-law distributed networks
of large size the transition is asymptotically discontinuous, 
if cooperativity is sufficiently high; however strong
size effects may conceal the real nature of the transition in
finite systems, so that the transition appears continuous even
for large networks.
We then present an analytical heterogeneous mean-field approach 
to the problem, which allows us to derive a number of predictions,
including the position of the threshold, the nature of the transition
and the associated critical exponent.
Numerical simulations validate the analytical predictions.
The theoretical approach indicates that for scale-free networks, 
with diverging second moment of the degree distribution, the transition
can only be continuous.

The paper is organized as follows. 
In Sec.~\ref{sec:model} we describe the CGCG model in detail.
Sec.~\ref{sec:results1} is devoted to a numerical investigation
of the nature of the transition for power-law degree-distributions,
showing how finite size effects can hinder the emergence of the
discontinuity in scale-rich networks.
Sec.~\ref{sec:theory} is devoted instead to the
theoretical analysis of the co-infection dynamics by means of a
heterogeneous mean--field approach. 
Theoretical results are compared with numerical simulations,
revealing a satisfactory agreement.
We present conclusions and an outlook in Sec.~\ref{sec:conclusion}.

\section{Model}
\label{sec:model}
In the classical SIR model in discrete time, each individual can be in
one of three different states: susceptible (S, individuals that are
healthy, neither infected nor immune), infected (I, individuals who can
transmit the disease), or removed (R, dead or recovered and immunized
individuals). At each time step each infected individual spontaneously
decays with probability $r$ into the removed state, while she
transmits the infection to each of her susceptible neighbors with
probability $p$.

The CGCG model is a modification of the SIR model with two 
circulating diseases, A and B.  
The infection probability for one disease is increased if the
individual currently has, or has had in the past, the other disease: 
Individuals uninfected by either disease get infected (with either 
A or B) by any infective neighbour with probability $p$, while a node 
that is or has been infected by one disease has a higher probability 
$q>p$ to get infected by the other pathogen.  
When recovering from one disease an individual becomes immune to that 
disease, but she can still be infected by the other.
The model is totally symmetric with respect to A and B.
Since each individual can be in one of three possible states (S, I, R)
with respect to each of the two diseases (A, B) there are nine
possible states for each individual, denoted by S, A, B, AB, a, b, aB,
Ab and ab, where, for each disease, capital letters refer to the infected
state, while lower--case letters refer to the removed state. 
States denoted by single letters (a, b, A, B) refer to states where the 
individual is still susceptible with respect to the other disease.

We simulate the CGCG model on power-law distributed networks 
$P(k) \sim k^{-\gamma}$, generated according to the uncorrelated
configuration model~\cite{Catanzaro2005}, for various values of the
$\gamma$ exponent.
We fix the recovery probability to $r=1$ throughout the paper.
In the various simulations, a number  $N_r$ (ranging from 200 to 2000) 
of independent realizations are performed.
We consider an initial condition
with all individuals initially in the susceptible (S) state
except for a single, randomly chosen individual who is in the 
doubly-infected AB state.

\section{Finite size effects and the nature of the transition for high cooperativity}
\label{sec:results1}

We start by analyzing the behavior of the CGCG model on
power-law degree-distributed
networks.  In Ref.~\cite{Chen2015} only continuous transitions were observed 
for the Barabasi--Albert (BA) scale--free network.
This is in contrast with the behavior of the same model
on Erd{\"o}s--R{\'e}nyi networks, which exhibits a discontinuous
transition for sufficiently high cooperativity.
One may wonder whether the continuity of the transition for BA networks
persists in the infinite size limit and whether the same occurs for other
broadly distributed networks, which are however not scale-free.

\begin{figure}
\includegraphics[width=.48\textwidth]{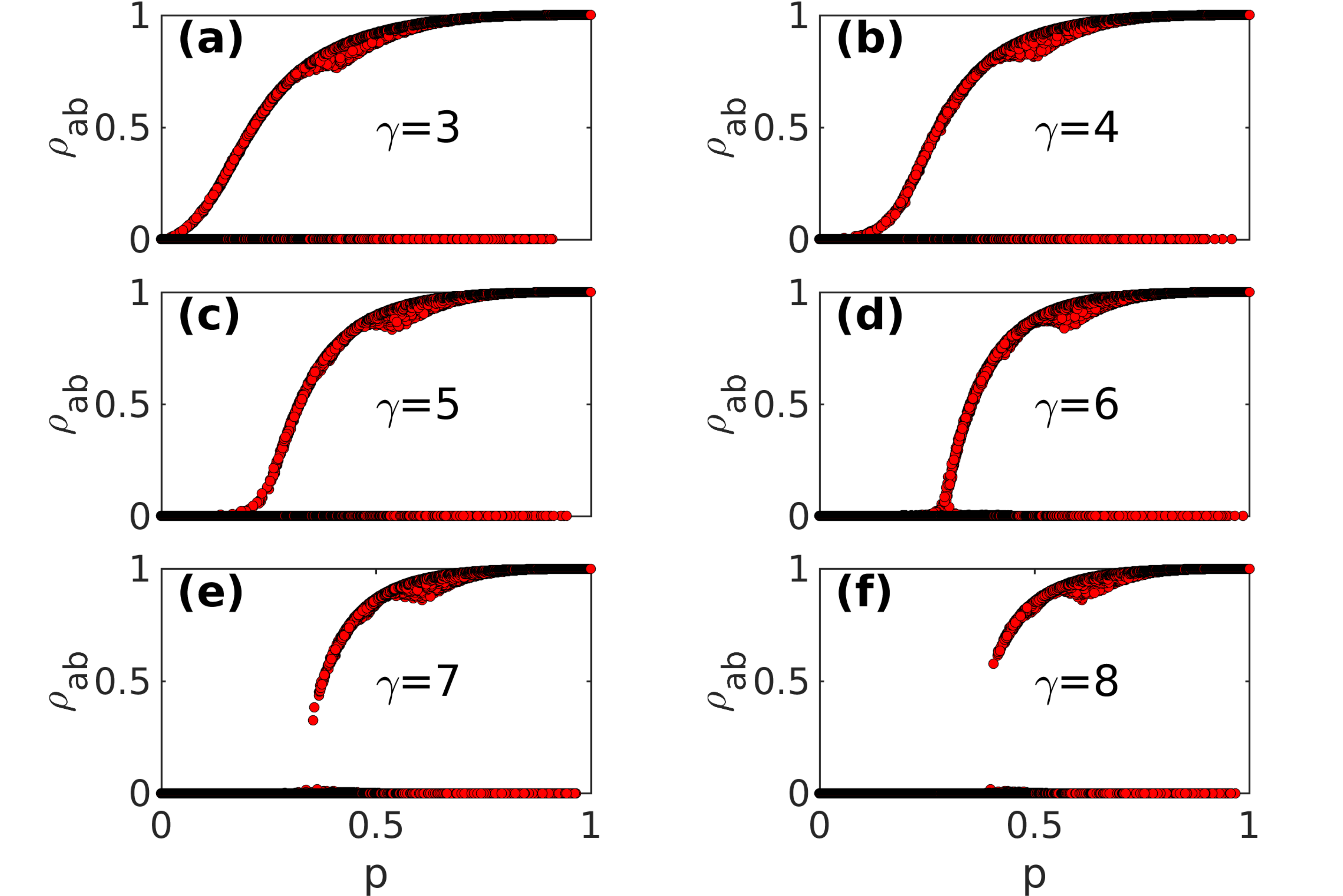}
\caption{The six panels represent the total density $\rho_{ab}$ of population
  recovered from both A and B in the final state plotted versus
  $p$ on scale--free networks with $\gamma=3,4,5,6,7,8$, built by
  applying the uncorrelated configuration model. 
  The system size is $N=10^5$, and $q=0.99$. 
  The results are averaged over a number $N_r$ of independent realizations
  that is $200$ for $\gamma=3$ to 5 and is increased to $2000$ 
  for $\gamma=6$ to 8.}
\label{fig:sf345678}
\end{figure}

The interpretation for the existence of a discontinuous transition put forward 
in Ref.~\cite{Chen2015} provides some hint on the issue.
The origin of the discontinuity is traced back to the relative abundance of 
long loops compared to the scarcity of short loops in the topology.
It is well known~\cite{Newman2002} that the clustering coefficient of 
uncorrelated random networks is
\be
C=\frac{\av{k}}{N} \left(\frac{\av{k^2}-\av{k}}{\av{k}^2} \right)^2.
\label{C}
\ee
Hence the clustering coefficient vanishes for large size for any $\gamma>2$.
However, this occurs differently depending on the value of $\gamma$.
The factor in parentheses in Eq.~(\ref{C}) is finite for $\gamma>3$, but
its value becomes larger as $\gamma$ gets smaller.
Hence we do expect that for systems of fixed size $N$ the transition
will be discontinuous for large $\gamma$, but the residual clustering
will make the transition appear continuous as $\gamma$ is reduced.
This expectation is tested by simulating the CGCG dynamics on networks
of fixed size ($N=10^5$), large cooperativity ($q=0.99$) 
and growing values of $\gamma$ starting from 
$\gamma=3$ (Fig.~\ref{fig:sf345678}).
As expected, upon increasing $\gamma$, the transition becomes 
discontinuous: short loops become less abundant as $\gamma$ grows
so that the two infections evolve separately and give rise to 
cooperative effects only after they meet following one long loop in 
the network.

\begin{figure}
\includegraphics[width=.48\textwidth]{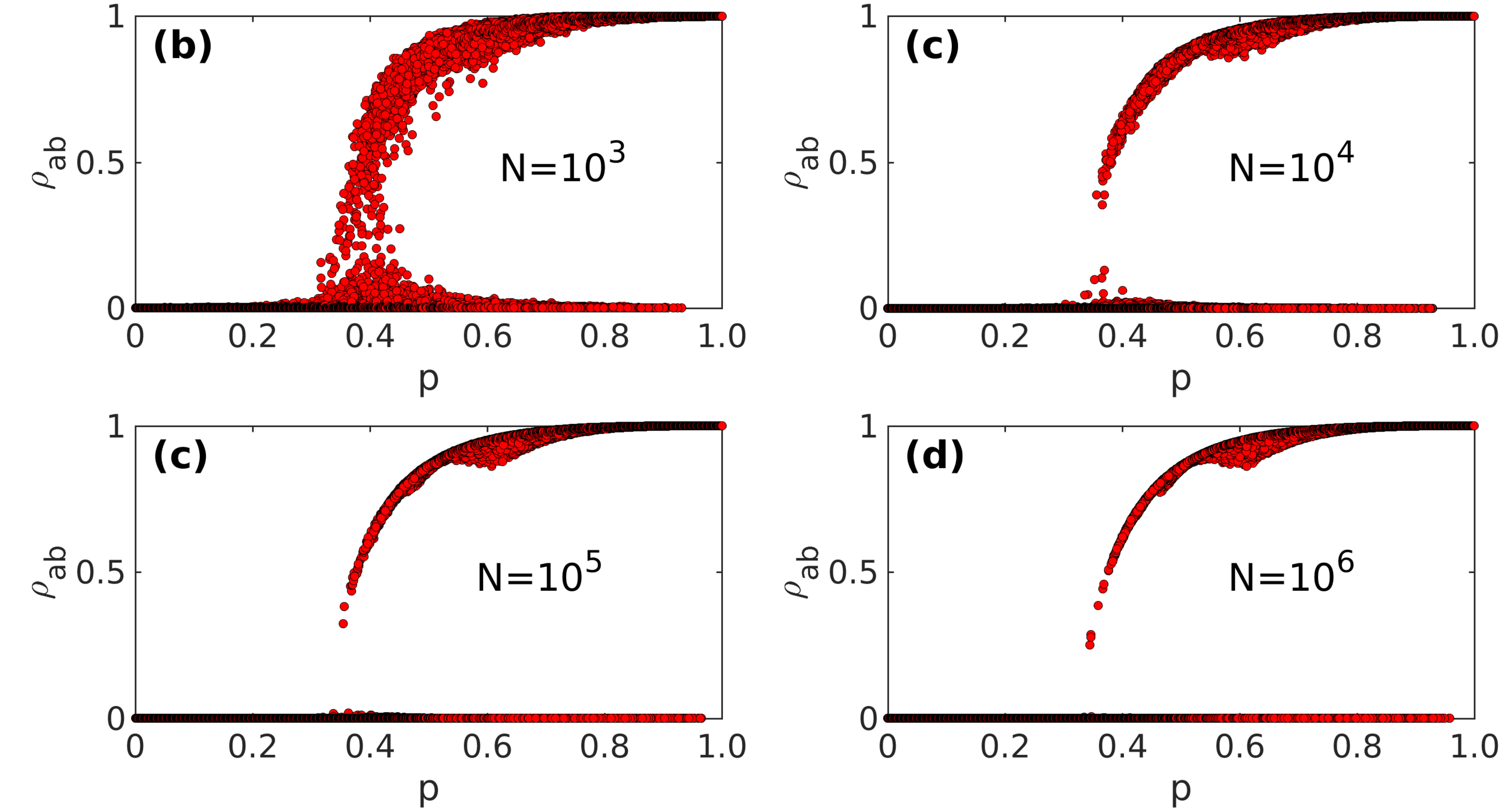}
\caption{Total density $\rho_{ab}$ of population
  recovered from both A and B in the final state, plotted versus
  $p$ on scale--free networks of different sizes
  ($N=10^3,10^4,10^5,10^6$, respectively), with $\gamma=7$. 
  The parameter $q$ is $0.99$.
  The results are averaged over $N_r=200,500,2000,2000$
  independent realizations in panels a,b,c,d, respectively.}
\label{fig:sfdifsize}
\end{figure}

Further evidence about the same effect is obtained by simulating 
the CGCG dynamics on a network with fixed $\gamma=7$ and increasing 
system size from $N=10^3$ to $10^6$ (see Fig.~\ref{fig:sfdifsize}): 
when the size is big enough the discontinuity appears.  

\begin{figure*}
\includegraphics[width=.98\textwidth]{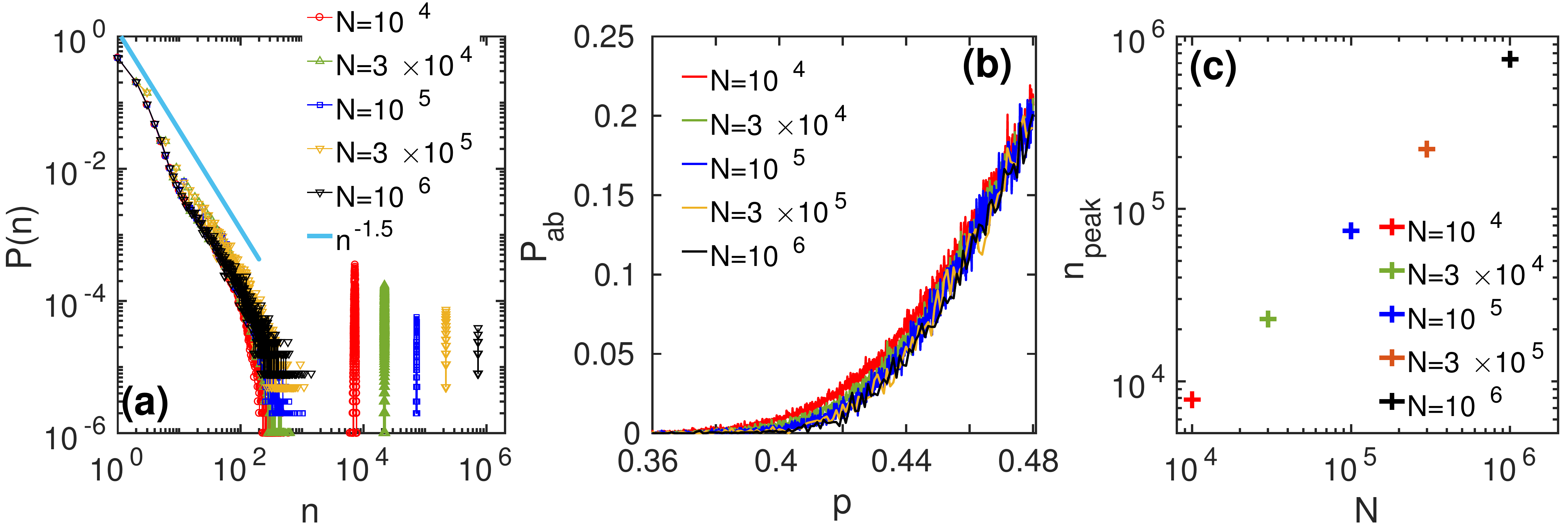}
\caption{
(a) Mass distribution of doubly-infected clusters at the threshold
$p_c= \av{k}/(\av{k^2} -\av{k}) = 0.43388$. 
The number of realizations is at least $N_r=10^5$
and various values of $N$. 
(b) Probability $P_{ab}$ of forming a giant cluster (with size larger 
than $0.05 N$) as a function of $p$ and system sizes ranging (top to bottom) 
from $N=10^4$ to $N=10^6$.
The number of realizations is $N_r=2000$.
(c) Plot of the size of the giant cluster as a function of $N$ for $p=p_c$,
showing a linear growth. All data are for $\gamma=7$ and $q=0.99$.}
\label{fig:hybrid}
\end{figure*}

The transition presented in Fig.~\ref{fig:sfdifsize} is discontinuous
but Fig.~\ref{fig:hybrid} shows that it is in fact of hybrid type, as
discussed in Ref.~\cite{Chen2015}.
This means that already at the transition point there may be
giant infected clusters occupying a finite fraction of the system
(see Fig.~\ref{fig:hybrid}(a) and ~\ref{fig:hybrid}(c)); this is at
odds with what happens in the usual, continuous, percolation transition.
However the probability that one of these clusters appears undergoes
a continuous transition at the threshold (Fig.~\ref{fig:hybrid}(b))
as in normal percolation.

For values $2<\gamma<3$, Eq.~(\ref{C}) points out that, 
since $\av{k^2}$ diverges as $k_{max}^{2(3-\gamma)}=N^{(3-\gamma)}$, 
the decay of $C$ with the system size becomes $N^{2-\gamma}$, slower than $1/N$.
This qualitative change does not allow to draw firm conclusions on the nature
of the transition for $\gamma<3$. Asymptotically the clustering coefficient
vanishes also in this case, but it does so slowly so that it might be
impossible for the two epidemics to grow initially separated clusters. 
Therefore we cannot say whether the continuous transition observed 
in Ref.~\cite{Chen2015} is a finite size effect or not. 
More on this issue will be provided by the analytical results in the 
next Section.

\section{Heterogeneous Mean-Field Theory for uncorrelated networks}
\label{sec:theory}
In this section we present an analytical approach to the cooperative
dynamics on uncorrelated networks, based on the heterogeneous 
mean-field (HMF) theory~\cite{PastorSatorras2001,Moreno2002}.
It is a refinement of the homogeneous mean-field used in 
Ref.~\cite{Chen2013}, that allows for quantities to depend on 
the degree $k$ of the node considered and is thus suitable also 
for topologies with a broad connectivity distribution.
For example, $[A]_k$ is the fraction of individuals in state A, restricted 
to nodes of degree $k$.

Following the same formalism of~\cite{Chen2013}, and exploiting the
full symmetry among A and B pathogens, we define the quantities
\begin{eqnarray}
Y_k & = & [A]_k+[a]_k =  [B]_k+[b]_k \\  
X_k & = & [A]_k+[Ab]_k+[AB]_k  =  [B]_k+[aB]_k+[AB]_k \\
R_k & =& [b]_k+[bA]_k+[ab]_k=[a]_k+[aB]_k+[ab]_k \,.
\label{eq:symmetry}
\end{eqnarray}
$Y_k$ is the fraction of nodes of degree $k$ which have contracted 
one of the infections and are susceptible with respect to the other; 
$X_k$ is the fraction of nodes of degree $k$ which are currently 
infected by one of the pathogens, regardless of the state with respect
to the other infection;
$R_k$ is instead the fraction of nodes which have recovered from
one infection, regardless of the state with respect to the other.
Eqs.~(4) of Ref.~\cite{Chen2013} can thus be generalized to
\begin{subequations}
\begin{align}
\dot{S}_{k} & = -2\alpha k S_{k} \Theta_X \\ \label{eq:egep1}
\dot{Y}_{k} & =  k(\alpha S_{k} - \beta Y_{k}) \Theta_X \\  \label{eq:egep2}
\dot{X}_{k} & =  k(\alpha S_{k} + \beta Y_{k}) \Theta_X -\mu X_k \\  \label{eq:egep3} 
\dot{R}_{k} & =  \mu X_{k} \,,
\end{align}
\label{eq:egep4}
\end{subequations}
which conserve the total probability $S_k+Y_k+X_k+R_k=1$.
The rates for the transmission of the first infection, of the second infection 
and for recovery are $\alpha$, $\beta$ and $\mu$, respectively.
$\Theta_X$ is the probability that any given edge points to
an infected node and is capable of transmitting the disease.
For SIR dynamics one must take into account the fact
that an infected individual cannot transmit the infection 
through the edge from which she was infected, hence~\cite{Boguna2003}
\be
\Theta_X = \frac{1}{\av{k}} \sum \limits_{k}(k-1) P(k) X_{k}(t).
\ee
In the following we set the recovery rate $\mu=1$,
with no loss of generality. 
Notice that Eqs.~(\ref{eq:egep4}) are for continuous dynamics, while
the dynamics considered in simulations are discrete, with lifetime of
the infected state deterministically equal to 1. 
In order to compare results between theory and simulation, we must
therefore consider the mapping between the rates and the
probabilities $\alpha=p/(1-p)$ and $\beta=q/(1-q)$ (see Appendix A).

Considering the initial condition with only one doubly-infected node 
$S_k(0) \simeq 1$, $R_k(0)=Y_k(0)=0$ and $X_k(0) = 1-S_k(0) \simeq 0$, 
Eqs.~\ref{eq:egep4} are readily integrated, yielding
\begin{subequations} 
\begin{align}
S_{k}(t) & =  e^{-2\alpha k\phi(t)} \\ \label{eq:stime1}
Y_{k}(t) & =  \frac{\alpha}{\beta-2\alpha} 
             [e^{-2\alpha k\phi(t)}-e^{-\beta k \phi(t)}] \\ 
R_{k}(t) & =  \int_{0}^{t} X_{k}(\tau)d \tau \label{eq:stime2},
\end{align}
\label{eq:stime}
\end{subequations}
where the auxiliary function $\phi(t)$ is defined as

\be
\label{eq:auxiliary}
\phi(t)  =  \int_{0}^{t}d\tau \Theta_X(\tau) 
	     =  \frac{1}{\av{k}} \sum_{k} (k-1) P(k) R_k(t).
\ee 

By deriving Eq.~(\ref{eq:auxiliary}) with respect to time we obtain
\begin{eqnarray}
\dot{\phi}(t) & = &\frac{1}{\av{k}} \sum_k (k-1) P(k) X_k \\  
	      & = &\frac{1}{\av{k}} \sum_k (k-1) P(k) [1-R_k-S_k-Y_k] \\ \nonumber
              & = & 1-\frac{1}{\av{k}} - \phi(t) \\ \nonumber
              & - & \frac{1}{\av{k}}
                 \left(1+\frac{\alpha}{\beta-2\alpha}\right) \sum_k(k-1) P(k)
                 e^{-2\alpha k\phi(t)} \\ 
              & + &   
                \frac{1}{\av{k}} \frac{\alpha}{\beta-2\alpha} 
                \sum_k (k-1)P(k) e^{-\beta k\phi(t)}\,.
\label{eq:timederivation}
\end{eqnarray}

At the end of the spreading process $X_k(\infty)=0$ and
$\lim_{t\rightarrow \infty} \dot{\phi}(t)=0$.
Hence $\phi_{\infty}$, the asymptotic value of $\phi(t)$ obeys
\begin{eqnarray}
\label{eq:expressionphi}
\phi_{\infty}  & = & 1-\frac{1}{\av{k}} \\ \nonumber
                       & - & \frac{1}{\av{k}}
              \left(1+\frac{\alpha}{\beta-2\alpha}\right) 
               \sum_k(k-1) P(k) e^{-2\alpha k \phi_{\infty}} \\ \nonumber
              & + &   
              \frac{1}{\av{k}} \frac{\alpha}{\beta-2\alpha} 
              \sum_k (k-1)P(k) e^{-\beta k\phi_{\infty}}\,.
\end{eqnarray}

Besides the zero solution $\phi_{\infty}=0$, a non-zero solution
representing endemic outbreaks can be obtained only
if the derivative of the r.h.s. of Eq.~(\ref{eq:expressionphi}) 
with respect to $\phi_{\infty}$, evaluated for $\phi_{\infty}=0$,
is larger than 1, i.e.,
\begin{eqnarray}
\frac{\alpha}{\av{k}}\sum_k k(k-1)P(k) \geq 1.
\label{eq:finalcondition}
\end{eqnarray} 
Consequently, the epidemic threshold is
\begin{eqnarray}
\alpha_c=\frac{\av{k}}{\av{k^2} -\av{k}} \,.
\label{eq:threshold}
\end{eqnarray}
Notice that the threshold does not depend on $\beta$
and is equal to the threshold for the SIR spreading of a single disease.
\begin{figure*}
\includegraphics[width=.98\textwidth]{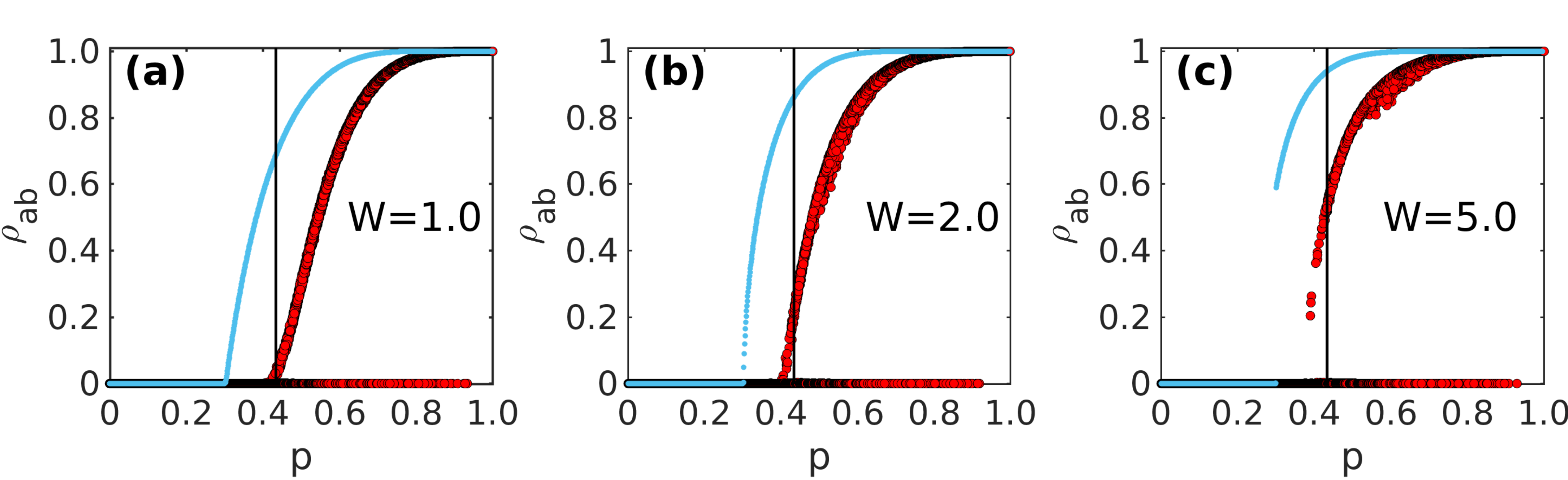}
\caption{Red large symbols represent the final fraction $\rho_{ab}$ of 
doubly-infected nodes on power-law networks of size $N=10^5$ with 
$\gamma=7$. They are plotted vs $p$
for various fixed values of the cooperativity $W=\beta/\alpha$.
The blue small symbols are the results of the numerical integration of
Eq.~(\ref{eq:egep4}). The plotted values are computed as
$\rho_{ab}=\sum_k [R_k(\infty)-Y_k(\infty)]$.
The black vertical line is $\av{k}/(\av{k^2}-\av{k})$.
}
\label{variousW}
\end{figure*}
This conclusion is supported by the simulation results
illustrated in Fig.~\ref{variousW}.
As the cooperativity $W=\beta/\alpha$ grows, the nature of the
transition passes from continuous to discontinuous; however, 
the position of the threshold remains unchanged.
Notice that the independence of the threshold value from $W$ is another
piece of evidence in favor of the interpretation of the transition
put forward in~\cite{Chen2015}. If the discontinuous 
transition is originated by the encounter of macroscopic singly-infected
clusters, such event is possible only when $p$ is larger than the
single-infection threshold. This is what is observed in Fig.~\ref{variousW}.

In the same figure we also plot the results of the numerical
integration of the HMF Eqs.~(\ref{eq:egep4}).
The agreement between the theoretical and numerical results is
only qualitative: the nature of the transition is the same, but the
analytical approach does not predict precisely the position of the
threshold. This is not a surprise, as it is known that the HMF theory
does not give accurate estimates of the SIR threshold for homogeneous
networks (see Appendix A). 

For single infection SIR the critical value $\alpha_c$ predicted by the HMF 
for continuous time dynamics has exactly the same value
of the exact threshold $p_c$ of discrete time dynamics.
The same seems to occur for SIR co--infections:
the threshold $p_c$ found in numerical simulations of the discrete 
dynamics is,
again by coincidence,  very close to the value $\alpha_c$ predicted by
HMF theory, and given by Eq.~(\ref{eq:threshold}).

The analytical approach allows also to derive the conditions under which 
the transition is continuous and the
associated critical exponent $1/\psi$, defined as 
\be
\phi_{\infty} \sim \left( \frac{\alpha-\alpha_c}{\alpha_c} \right)^{1/\psi}.
\ee

\begin{figure}
\includegraphics[width=.48\textwidth]{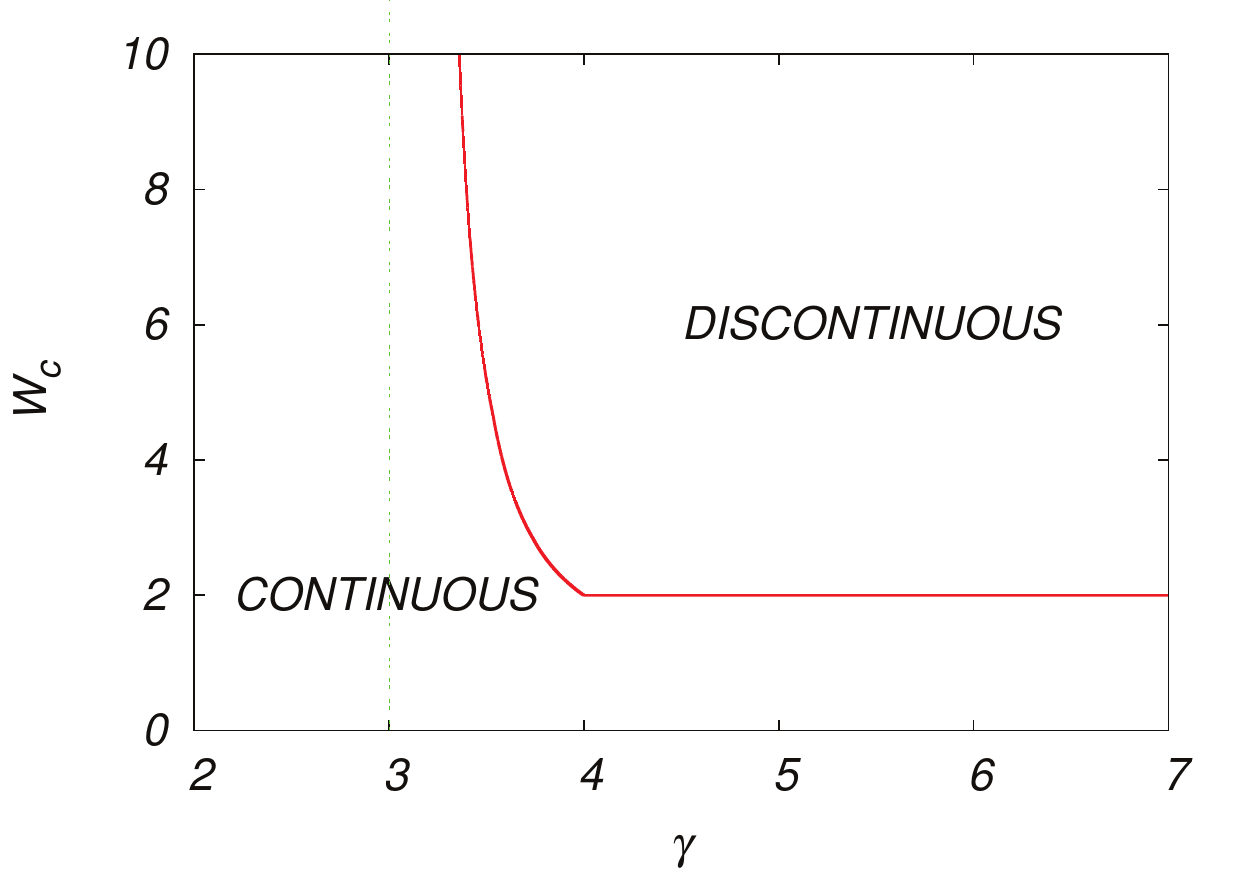}
\caption{The value of the minimum cooperativity $W_c$ needed for
a discontinuous transition, as a function of $\gamma$.}
\label{figW}
\end{figure}

The detailed calculations, reported in Appendix B, give the
following results, which provide a complete picture of the model
behavior:
\begin{itemize}
\item 
For $\gamma>4$ the transition is continuous if the cooperativity 
$W=\beta/\alpha$ is sufficiently small, i.e., $W<W_c=2$ 
and the associated exponent is
$1/\psi=1$. When $W>W_c$ instead the transition is discontinuous.
This is in agreement with what is found in Fig.~\ref{variousW}.
In the marginal case $W=W_c$, the transition is still continuous
but with critical exponent $1/(\gamma-3)$ for $4<\gamma<5$
and $1/2$ for $\gamma>5$. This last result is consistent with the
mean-field treatment in Ref.~\cite{Chen2013}.

\item
For $3<\gamma<4$ the limit value $W_c$ for the transition
to be continuous is always larger than $2$ and assumes a nontrivial
dependence on $\gamma$ (see Fig.~\ref{figW}) that continuously 
interpolates between $2$ for $\gamma \to 4$ and infinity for $\gamma \to 3$. 
Also the critical exponent changes continuously
$1/\psi = 1/(\gamma-3)$. Also in this range of $\gamma$ values 
things are different for marginal cooperativity $W=W_c$: 
in this case $1/\psi=1$ regardless of the value of $\gamma$.

\item
For $2<\gamma<3$ the threshold vanishes, the transition is 
continuous for any value of the cooperativity $W$ and the growth 
of the order parameter for small $\alpha$ is $\alpha^{1/\psi}$, where
$1/\psi = (\gamma-2)/(3-\gamma)$.

\end{itemize}

\section{Conclusions}
\label{sec:conclusion}

In this paper we have performed an analysis of the implications
of cooperativity for SIR-like epidemic spreading on power-law
degree-distributed networks.
First, we have tested the interpretation put forward
in Ref.~\cite{Chen2015} about the physical origin of the discontinuous 
epidemic transition occurring in this kind of systems for large
cooperativity.
Via numerical simulations, we have shown that strong finite system
size effects may hide the discontinuity, but that if the system size 
is large enough the true, hybrid, nature of the transition becomes evident.
This provides a strong validation of the physical picture proposed by
Cai et al.~\cite{Chen2015}.

In the second part of the paper we have applied the 
heterogeneous mean-field theoretical approach to the co--infection 
dynamics on uncorrelated networks. In this way we have derived
an expression for the epidemic threshold, which does not depend
on cooperativity, in agreement with simulations. 
Moreover, the approach allows us to derive, for power-law 
degree-distributed topologies,
the range of cooperativity values for which the transition
is continuous and the associated critical exponent.
We find a nice confirmation that the discontinuous
transition occurs when cooperativity is sufficiently high and
local clustering is instead small.
An interesting point in this respect regards what
happens for scale-free networks, i.e. for $2<\gamma<3$.
In such a case HMF theory predicts that the transition is 
continuous for any value of the cooperativity $W$. 
This is in agreement with the simulation results for BA
networks presented in Ref.~\cite{Chen2015,Chen2016} for
networks of moderate size. 
However, one cannot in principle exclude that also for
$\gamma<3$ the continuous transition observed in simulations
is just a finite size effect. While HMF theory seems to indicate
the contrary, the global clustering coefficient Eq.~(\ref{C})
vanishes as the system size diverges also for $\gamma<3$, so that
asymptotically there are no short loops in the system.
Numerical simulations cannot be conclusive about this question.
The issue could be clarified only by a careful analysis of the
balance between the relative abundance of short loops (encoded by $C$)
and long loops in the network. It remains as a challenging goal.

It is also worth noticing that our theoretical approach is 
fully general with respect to the value of the infection 
rates ratio $W=\beta/\alpha$. Therefore our results provide
predictions also for competing pathogens, for which $W<1$. 
In that case our theory predicts always a continuous transition,
for any $\gamma$ perfectly analogue to the case of weak cooperativity 
$1<W<2$.

Mutual cooperation is a very important ingredient also for
non-biological spreading processes, such as the diffusion of 
ideas or the adoption of innovations.
The investigation of cooperative effects in general complex
contagion dynamics remains a very interesting avenue for future 
research activity.

\section*{Acknowledgments}
This work was supported by China Postdoctoral Science Foundation No.
2015M582532 and by China Scholarship Council (CSC), grant
number 201606075021.

\section*{Appendix A}
In this Appendix, we summarize the connection between bond percolation
and the various implementations of the SIR dynamics to allow a proper
comparison between theoretical and numerical results.
We refer to single-infection SIR in this Appendix.

\subsection*{Bond percolation}
In bond percolation on networks, each edge is occupied
with probability $T$.
The generating function formalism~\cite{Callaway2000} provides a 
solution for bond percolation that is exact for random uncorrelated 
networks with any degree distribution $P(k)$ in the limit of infinite 
size $N \to \infty$~\cite{Dorogovtsev2008}
\be
T_c = \frac{\av{k}}{\av{k^2}-\av{k}}.
\label{eq0}
\ee
Notice that for Erd\"os-R\'enyi graphs $\av{k^2}=\av{k}^2+\av{k}$, hence
$T_c=1/\av{k}$. For Random Regular Graphs of degree $k$: $T_c=1/(k-1)$.

\subsection*{SIR in discrete time}

Consider the following implementation of the SIR model: at time step $t$
one goes through all nodes in state I. 
Each neighbor in state S of each I node is infected with probability $p$. 
When all infected nodes at time step $t$ have been considered, they all
go to state R and $t \to t+1$. This model is also called Independent
Cascade Model.

The lifetime $\tau$ of each node in the infected state I is fixed 
deterministically $\tau=1$. With this condition it is possible to 
exactly map the static properties of the model to bond
percolation~\cite{Newman2002b}.
One defines as transmissibility the probability for an infected node
to pass the infection along one edge before recovering.
In the present model this transmissibility is trivially $T=p$.
The epidemic process is perfectly equivalent to a bond percolation
process starting from a node and iteratively occupying neighbors with 
transmissibility $T=p$.
Therefore the epidemic threshold is
\be
p_c = T_c = \frac{\av{k}}{\av{k^2}-\av{k}}.
\label{eqa1}
\ee
\subsection*{SIR in continuous time}

It is possible to define SIR dynamics also in continuous time.
Each infected node has a rate (probability per time unit) $\mu$ to recover.
Each edge connecting a node I to a node S has a rate
$\alpha$ of transmitting the infection.
In this model the lifetime $\tau$ of the infected state is a stochastic
variable, distributed as $P(\tau) = \mu \exp(-\mu \tau)$, so that
$\av{\tau}=1/\mu$.
For this reason, it is not possible to map exactly the model onto
bond percolation. However, it is possible to perform an approximate
mapping~\cite{Newman2002b}, which turns out to be exact in the determination
of the epidemic threshold~\cite{PastorSatorras2015}.

The idea is to compute the average transmissibility $\av{T}$, i.e.
the average probability to transmit the infection before recovering.
For a given $\tau$ the probability of not transmitting the infection 
during time $\tau$ is $1-T = \exp(-\alpha \tau)$.
Therefore the average transmissibility is
\be
\av{T} = 1 - \int_0^\infty d\tau P(\tau) \exp(-\alpha \tau).
\ee
Inserting the expression for $P(\tau)$ one gets
\be
\av{T} = \frac{1}{\mu/\alpha+1}.
\ee
The transition will occur when $\av{T}$ equals the expression in 
Eq.~(\ref{eqa1}), yielding
\be
\left(\frac{\alpha}{\mu} \right)_c =  \frac{\av{k}}{\av{k^2}-2\av{k}}.
\label{eqa2}
\ee

Notice that this expression is {\em different} from the expression for
discrete time SIR. Both are exact (in the limit of infinite size)
but for different types of dynamics.
 
In the case of Erd\H{o}s-R\'enyi graphs the previous formula implies
$(\alpha/\mu)_c =  \frac{1}{\av{k}-1}$. For Random Regular Graphs
instead
$(\alpha/\mu)_c =  \frac{1}{k-2}$.

A consequence of this treatment is that one can pass from the
parameter $p$ of the discrete dynamics to the parameter $\alpha/\mu$
of the continuous dynamics (the only relevant parameter) by the relations
\be
p = \frac{1}{\mu/\alpha+1}~~~~~~~~~\frac{\alpha}{\mu} = \frac{p}{1-p}.
\ee

\subsection*{HMF theory for SIR}
The SIR model in the continuous time formulation can be attacked by
means of the Heterogeneous Mean-Field (HMF) theory.
The HMF theory allows to derive an approximate formula for the epidemic 
threshold~\cite{PastorSatorras2015}
\be
\left(\frac{\alpha}{\mu} \right)_c^{HMF} =  \frac{\av{k}}{\av{k^2}-\av{k}}.
\ee
It is very important to stress that:
\begin{itemize}
\item 
The HMF prediction is {\em not exact}.
The exact result for continuous dynamics is Eq.~(\ref{eqa2}).
This discrepancy is not surprising, as transition points are
not universal, they depend on details of the dynamics and are
rarely predicted with accuracy by approximate mean-field approaches.

\item
The HMF prediction for the parameter $\alpha$ in continuous SIR 
dynamics coincides with the exact threshold for the parameter $p$
of discrete SIR dynamics (Eq.~\ref{eqa1}). 
This seems to be just a coincidence.
\end{itemize}

\section*{Appendix B}
\label{sec:appendix}

In this Appendix, we consider the expansion of 
Eq.~(\ref{eq:expressionphi}) for small $\phi_{\infty}$ and, based on
it, we deduce the properties of the transition: its position $\alpha_c$,
its nature and the associated critical exponent $1/\psi$.

To simplify notation, we write $\phi$ for the stationary
value $\phi_{\infty}$ and $m$ for the minimum degree $k_{min}$.
Introducing the cooperativity $W=\beta/\alpha$, 
Eq.~(\ref{eq:expressionphi}) can be rewritten as
\be
\phi=\frac{\langle k-1\rangle}{\langle k \rangle}
+\frac{1}{\av{k}} \sum_k(k-1)P(k)
\frac{e^{-W\alpha k\phi}-(W-1)e^{-2\alpha k \phi}}{W-2}
\label{selfcons01}
\ee

By assuming the explicit form of the degree distribution 
$P(k)=(\gamma-1) m^{\gamma-1} k^{-\gamma}$ and transforming the sum
in a continuous integral, it is possible to rewrite Eq.~\ref{selfcons01}
\begin{equation}
\begin{array}{lll}
\phi=
\frac{\langle k-1\rangle}{\langle k\rangle }\!\!\!
&
-\frac{\gamma-1}{\langle k \rangle}
\frac{W-1}{W-2}\left[\right.
&\!\!
m(2\alpha m\phi)^{\gamma-2}\Gamma(2-\gamma,2\alpha m \phi)
\\
&&
\left.\!
-(2\alpha m\phi)^{\gamma-1}\Gamma(1-\gamma,2\alpha m \phi) 
\right]
\\
&
+\frac{\gamma-1}{\langle k \rangle}
\frac{1}{W-2}\left[\right.
&\!\!
m(W\alpha m \phi)^{\gamma-2}\Gamma(2-\gamma,W\alpha m\phi)
\\
&&\left.\!
-(W\alpha m \phi)^{\gamma-1}\Gamma(1-\gamma,W\alpha m\phi) 
\right]
\end{array}
\label{selfcons1}
\end{equation}
where $\Gamma(s,x)=\int_x^{\infty}t^{s-1}e^{-t}dt$ is the incomplete 
Gamma function, that can be expanded as
\begin{equation}
\Gamma(s,x)=\Gamma(s)-x^s\sum_{n=0}^{\infty}
\frac{(-x)^n}{n!(s+n)}\,.
\label{gamma}
\end{equation}

Replacing the expansion in Eq.~(\ref{selfcons1}) gives
\begin{widetext}
\begin{equation}
\begin{array}{ll} 
\phi=\frac{\langle k-1\rangle}{\langle k\rangle }\!\!
&
-\frac{(\gamma-1)}{\langle k \rangle}
\frac{W-1}{W-2}
\left[
m\left(
(2m\alpha\phi)^{\gamma-2}\Gamma(2-\gamma)
-\sum_{n=0}^{\infty}\frac{(-2\alpha m \phi)^n}{(2-\gamma+n)n!}
\right)
-
\left(
(2m\alpha\phi)^{\gamma-1}\Gamma(1-\gamma)
-\sum_{n=0}^{\infty}\frac{(-2\alpha m \phi)^n}{(1-\gamma+n)n!}
\right)
\right]+\\ 
&
+\frac{(\gamma-1)}{\langle k \rangle}
\frac{1}{W-2}
\left[m
\left(
(mW\alpha\phi)^{\gamma-2}\Gamma(2-\gamma)
-\sum_{n=0}^{\infty}\frac{(-W\alpha m \phi)^n}{(2-\gamma+n)n!}
\right)
-
\left(
(mW\alpha\phi)^{\gamma-1}\Gamma(1-\gamma)
-\sum_{n=0}^{\infty}\frac{(-W\alpha m \phi)^n}{(1-\gamma+n)n!}
\right)
\right]
\end{array}
\label{selfcons2}
\end{equation}
\end{widetext}

Collecting terms of the same order in $\phi$, 
and using the identity $\av{k}=m(\gamma-1)/(\gamma-2)$
we get
\begin{equation}
\begin{array}{ll}
\phi=\!\!&
\Gamma(3-\gamma) (2m\alpha)^{\gamma-2}
G(W,\gamma-2)\phi^{\gamma-2}+\\
&\!\!\!-\Gamma(2-\gamma)
\frac{(2m\alpha)^{\gamma-1}}{m(\gamma-1)/(\gamma-2)}
G(W,\gamma-1)\phi^{\gamma-1}+\\
&\!\!\!-\frac{\gamma-2}{m}
\sum_{n=1}^{\infty}\frac{(-2m\alpha)^n}{n!}
\left(\frac{m}{\gamma-(n+2)}-
\frac{1}{\gamma-(n+1)}\right)
G(W,n)\phi^n,
\end{array}
\label{selfcons4}
\end{equation}
where  we have defined  
\begin{equation}
G(W,n)=1-\frac{1}{2}\frac{(W/2)^{n}-1}{W/2-1}\,.
\end{equation}
Notice that all terms of zero-th order cancel out.

For $\gamma>4$ the lowest-order terms of the expansion are
\begin{equation}
\begin{array}{ll}
\phi=&\alpha
\left(
m\frac{\gamma-2}{\gamma-3}-1
\right)\phi+\\
&
+\frac{\gamma-2}{\gamma-3}
m\alpha^2(W-2)
\left(
m\frac{\gamma-3}{\gamma-4}-1
\right)\phi^2+O(\phi^2)
\label{selfcons5}
\end{array}
\end{equation}
The transition point $\alpha_c$ is the value for which the
linear part vanishes.
Therefore
\be
\alpha_c=\left(m\frac{\gamma-2}{\gamma-3}-1
\right)^{-1}\,\,\,\mbox{for}\,\,\,\,\gamma>4\,.
\label{alpha_c}
\ee
Notice that this expression is nothing else than Eq.~(\ref{eq:threshold}),
$\alpha_c=\av{k}/(\av{k^2} -\av{k})$.

In order for the transition to be continuous, 
Eq.~(\ref{selfcons5}) must admit a solution with arbitrary small 
$\phi$ for $\alpha$ slightly above the transition point
$(\alpha-\alpha_c)/\alpha_c=a\phi^\psi$, with $a>0$.
Inserting this expression into the r.h.s. of Eq.~(\ref{selfcons5})
we obtain
\begin{equation}
\phi=\phi(1+a\phi^\psi) +\frac{\gamma-2}{\gamma-3}
m(\alpha\phi)^2(W-2)
\left(m\frac{\gamma-3}{\gamma-4}-1
\right)
+O(\phi^2)
\label{selfcons6}
\end{equation}
that can be satisfied for $\psi=1$ as long as
\begin{equation} 
\frac{\gamma-2}{\gamma-3}
m\alpha^2_c(W-2)
\left(m\frac{\gamma-3}{\gamma-4}-1
\right)  <0.
\end{equation} 
This occurs for $W<W_c=2$, which means $\beta<2\alpha$. 
Therefore the transition becomes discontinuous for
high enough cooperativity, and precisely when the infectivity for the
second pathogen is at least twice the one for the single disease.

When the cooperativity $W$ exactly equals its critical value, the
coefficient of the term of order $\phi^2$ vanishes. 
The transition can still be continuous but the term $\phi^{1+\psi}$ has 
to match the next order term in the expansion.
For $4<\gamma<5$ such a term is $\phi^{\gamma-2}$ 
so that $\psi=\gamma-3$; for $\gamma>5$ the next order term is
$\phi^3$, implying $\psi=2$. 
In this last case the order parameter grows as $(\alpha-\alpha_c)^{1/2}$
in agreement with the mean-field results obtained in Ref.~\cite{Chen2013}.

For $3<\gamma<4$, the transition point is still given by Eq.~(\ref{alpha_c}).
Assuming again $(\alpha-\alpha_c)/\alpha_c=a\phi^\psi$ the lowest orders of 
Eq.~(\ref{selfcons4}) are
\begin{equation}
\phi=\phi(1+a\phi^\psi) +
\Gamma(3-\gamma) (2m\alpha\phi)^{\gamma-2}
G(W,\gamma-2)
+O(\phi^2)
\label{selfcons7}
\end{equation}
Eq.~(\ref{selfcons7}) can be satisfied if the term $a \phi^{1+\psi}$ 
matches the term of order $\phi^{\gamma-2}$. 
This is possible if $\psi=\gamma-3$ and if the coefficient 
$\Gamma(3-\gamma) (2m\alpha)^{\gamma-2} G(W,\gamma-2)$
is negative. Since for this range of $\gamma$, $\Gamma(3-\gamma)$
is negative, the condition for the transition to be continuous
is $G(W,\gamma-2)>0$. Therefore, for a given $3< \gamma <4$ the
transition is continuous only for $W \leq W_c$, with $W_c$ determined by
\begin{equation}
G(W_c,\gamma-2)=0\,,
\label{selfcons9}
\end{equation}
otherwise the transition is discontinuous.
Figure~\ref{figW} displays how $W_c$ depends on $\gamma$.
Starting from $\gamma=4$ and reducing $\gamma$ the value of the
cooperativity $W_c$ needed for the transition to become discontinuous
grows and it diverges as $\gamma \rightarrow 3$.
The critical exponent for the continuous transition is
$1/\psi=1/(\gamma-3)$ for $W<W_c$.
For $W$ exactly equal to $W_c$ the coefficient of the term of order 
$\phi^{\gamma-3}$ vanishes, therefore the continuous
transition is possible only when the term of order $\phi^{1+\psi}$ is
matched by the next term in the expansion, which for $3<\gamma<4$ is
given by the term in $\phi^2$, yielding $\psi=1$. 

Finally, let us consider the case $2<\gamma<3$. In this range the
threshold is $\alpha_c=0$ and we assume $\alpha = a' \phi^{\psi}$.
The lowest order of the r.h.s. of Eq.~(\ref{selfcons4}) is
$\Gamma(3-\gamma) (2m a' \phi^{\psi})^{\gamma-2} G(W,\gamma-2)\phi^{\gamma-2}$.
Since now $\Gamma(3-\gamma)>0$ and $G(W,\gamma-2)>0$ for any $W$ there
is always a solution (i.e., the transition is always continuous)
provided $(1+\psi)(\gamma-2)=1$, implying that
the critical exponent is $1/\psi=(\gamma-2)/(3-\gamma)$.

\bibliography{multireference}

\begin{thebibliography}{28}%
\makeatletter
\providecommand \@ifxundefined [1]{%
 \@ifx{#1\undefined}
}%
\providecommand \@ifnum [1]{%
 \ifnum #1\expandafter \@firstoftwo
 \else \expandafter \@secondoftwo
 \fi
}%
\providecommand \@ifx [1]{%
 \ifx #1\expandafter \@firstoftwo
 \else \expandafter \@secondoftwo
 \fi
}%
\providecommand \natexlab [1]{#1}%
\providecommand \enquote  [1]{``#1''}%
\providecommand \bibnamefont  [1]{#1}%
\providecommand \bibfnamefont [1]{#1}%
\providecommand \citenamefont [1]{#1}%
\providecommand \href@noop [0]{\@secondoftwo}%
\providecommand \href [0]{\begingroup \@sanitize@url \@href}%
\providecommand \@href[1]{\@@startlink{#1}\@@href}%
\providecommand \@@href[1]{\endgroup#1\@@endlink}%
\providecommand \@sanitize@url [0]{\catcode `\\12\catcode `\$12\catcode
  `\&12\catcode `\#12\catcode `\^12\catcode `\_12\catcode `\%12\relax}%
\providecommand \@@startlink[1]{}%
\providecommand \@@endlink[0]{}%
\providecommand \url  [0]{\begingroup\@sanitize@url \@url }%
\providecommand \@url [1]{\endgroup\@href {#1}{\urlprefix }}%
\providecommand \urlprefix  [0]{URL }%
\providecommand \Eprint [0]{\href }%
\providecommand \doibase [0]{http://dx.doi.org/}%
\providecommand \selectlanguage [0]{\@gobble}%
\providecommand \bibinfo  [0]{\@secondoftwo}%
\providecommand \bibfield  [0]{\@secondoftwo}%
\providecommand \translation [1]{[#1]}%
\providecommand \BibitemOpen [0]{}%
\providecommand \bibitemStop [0]{}%
\providecommand \bibitemNoStop [0]{.\EOS\space}%
\providecommand \EOS [0]{\spacefactor3000\relax}%
\providecommand \BibitemShut  [1]{\csname bibitem#1\endcsname}%
\let\auto@bib@innerbib\@empty
\bibitem [{\citenamefont {Pastor-Satorras}\ \emph {et~al.}(2015)\citenamefont
  {Pastor-Satorras}, \citenamefont {Castellano}, \citenamefont {Van~Mieghem},\
  and\ \citenamefont {Vespignani}}]{PastorSatorras2015}%
  \BibitemOpen
  \bibfield  {author} {\bibinfo {author} {\bibfnamefont {R.}~\bibnamefont
  {Pastor-Satorras}}, \bibinfo {author} {\bibfnamefont {C.}~\bibnamefont
  {Castellano}}, \bibinfo {author} {\bibfnamefont {P.}~\bibnamefont
  {Van~Mieghem}}, \ and\ \bibinfo {author} {\bibfnamefont {A.}~\bibnamefont
  {Vespignani}},\ }\href {\doibase 10.1103/RevModPhys.87.925} {\bibfield
  {journal} {\bibinfo  {journal} {Rev. Mod. Phys.}\ }\textbf {\bibinfo {volume}
  {87}},\ \bibinfo {pages} {925} (\bibinfo {year} {2015})}\BibitemShut
  {NoStop}%
\bibitem [{\citenamefont {Newman}(2005{\natexlab{a}})}]{Newman2005}%
  \BibitemOpen
  \bibfield  {author} {\bibinfo {author} {\bibfnamefont {M.~E.~J.}\
  \bibnamefont {Newman}},\ }\href {\doibase 10.1103/PhysRevLett.95.108701}
  {\bibfield  {journal} {\bibinfo  {journal} {Phys. Rev. Lett.}\ }\textbf
  {\bibinfo {volume} {95}},\ \bibinfo {pages} {108701} (\bibinfo {year}
  {2005}{\natexlab{a}})}\BibitemShut {NoStop}%
\bibitem [{\citenamefont {Funk}\ and\ \citenamefont {Jansen}(2010)}]{Funk2010}%
  \BibitemOpen
  \bibfield  {author} {\bibinfo {author} {\bibfnamefont {S.}~\bibnamefont
  {Funk}}\ and\ \bibinfo {author} {\bibfnamefont {V.~A.~A.}\ \bibnamefont
  {Jansen}},\ }\href {\doibase 10.1103/PhysRevE.81.036118} {\bibfield
  {journal} {\bibinfo  {journal} {Phys. Rev. E}\ }\textbf {\bibinfo {volume}
  {81}},\ \bibinfo {pages} {036118} (\bibinfo {year} {2010})}\BibitemShut
  {NoStop}%
\bibitem [{\citenamefont {Marceau}\ \emph {et~al.}(2011)\citenamefont
  {Marceau}, \citenamefont {No\"el}, \citenamefont {H\'ebert-Dufresne},
  \citenamefont {Allard},\ and\ \citenamefont {Dub\'e}}]{Marceau2011}%
  \BibitemOpen
  \bibfield  {author} {\bibinfo {author} {\bibfnamefont {V.}~\bibnamefont
  {Marceau}}, \bibinfo {author} {\bibfnamefont {P.-A.}\ \bibnamefont {No\"el}},
  \bibinfo {author} {\bibfnamefont {L.}~\bibnamefont {H\'ebert-Dufresne}},
  \bibinfo {author} {\bibfnamefont {A.}~\bibnamefont {Allard}}, \ and\ \bibinfo
  {author} {\bibfnamefont {L.~J.}\ \bibnamefont {Dub\'e}},\ }\href {\doibase
  10.1103/PhysRevE.84.026105} {\bibfield  {journal} {\bibinfo  {journal} {Phys.
  Rev. E}\ }\textbf {\bibinfo {volume} {84}},\ \bibinfo {pages} {026105}
  (\bibinfo {year} {2011})}\BibitemShut {NoStop}%
\bibitem [{\citenamefont {Miller}(2013)}]{Miller2013}%
  \BibitemOpen
  \bibfield  {author} {\bibinfo {author} {\bibfnamefont {J.~C.}\ \bibnamefont
  {Miller}},\ }\href {\doibase 10.1103/PhysRevE.87.060801} {\bibfield
  {journal} {\bibinfo  {journal} {Phys. Rev. E}\ }\textbf {\bibinfo {volume}
  {87}},\ \bibinfo {pages} {060801} (\bibinfo {year} {2013})}\BibitemShut
  {NoStop}%
\bibitem [{\citenamefont {Taubenberger}\ and\ \citenamefont
  {Morens}(2006)}]{Taubenberger2006}%
  \BibitemOpen
  \bibfield  {author} {\bibinfo {author} {\bibfnamefont {J.~K.}\ \bibnamefont
  {Taubenberger}}\ and\ \bibinfo {author} {\bibfnamefont {D.~M.}\ \bibnamefont
  {Morens}},\ }\href {\doibase 10.3201/eid1201.050979} {\bibfield  {journal}
  {\bibinfo  {journal} {Emerging Infectious Diseases}\ }\textbf {\bibinfo
  {volume} {12}},\ \bibinfo {pages} {15} (\bibinfo {year} {2006})}\BibitemShut
  {NoStop}%
\bibitem [{\citenamefont {Morens}\ \emph {et~al.}(2008)\citenamefont {Morens},
  \citenamefont {Taubenberger},\ and\ \citenamefont {Fauci}}]{Morens2008}%
  \BibitemOpen
  \bibfield  {author} {\bibinfo {author} {\bibfnamefont {D.~M.}\ \bibnamefont
  {Morens}}, \bibinfo {author} {\bibfnamefont {J.~K.}\ \bibnamefont
  {Taubenberger}}, \ and\ \bibinfo {author} {\bibfnamefont {A.~S.}\
  \bibnamefont {Fauci}},\ }\href {\doibase 10.1086/591708} {\bibfield
  {journal} {\bibinfo  {journal} {The Journal of Infectious Diseases}\ }\textbf
  {\bibinfo {volume} {198}},\ \bibinfo {pages} {962} (\bibinfo {year}
  {2008})}\BibitemShut {NoStop}%
\bibitem [{\citenamefont {Brundage}\ and\ \citenamefont
  {Shanks}(2008)}]{Brundage2008}%
  \BibitemOpen
  \bibfield  {author} {\bibinfo {author} {\bibfnamefont {J.~F.}\ \bibnamefont
  {Brundage}}\ and\ \bibinfo {author} {\bibfnamefont {G.}~\bibnamefont
  {Shanks}},\ }\href {\doibase https://dx.doi.org/10.3201/eid1408.071313}
  {\bibfield  {journal} {\bibinfo  {journal} {Emerg. Infect Dis.}\ }\textbf
  {\bibinfo {volume} {14}},\ \bibinfo {pages} {1193} (\bibinfo {year}
  {2008})}\BibitemShut {NoStop}%
\bibitem [{\citenamefont {Sulkowski}(2008)}]{Sulkowski2008}%
  \BibitemOpen
  \bibfield  {author} {\bibinfo {author} {\bibfnamefont {M.~S.}\ \bibnamefont
  {Sulkowski}},\ }\href {\doibase http://dx.doi.org/10.1016/j.jhep.2007.11.009}
  {\bibfield  {journal} {\bibinfo  {journal} {Journal of Hepatology}\ }\textbf
  {\bibinfo {volume} {48}},\ \bibinfo {pages} {353} (\bibinfo {year}
  {2008})}\BibitemShut {NoStop}%
\bibitem [{\citenamefont {Newman}\ and\ \citenamefont
  {Ferrario}(2013)}]{NewmanFerrario2013}%
  \BibitemOpen
  \bibfield  {author} {\bibinfo {author} {\bibfnamefont {M.~E.~J.}\
  \bibnamefont {Newman}}\ and\ \bibinfo {author} {\bibfnamefont {C.~R.}\
  \bibnamefont {Ferrario}},\ }\href {\doibase 10.1371/journal.pone.0071321}
  {\bibfield  {journal} {\bibinfo  {journal} {PLoS ONE}\ }\textbf {\bibinfo
  {volume} {8}},\ \bibinfo {pages} {e71321} (\bibinfo {year}
  {2013})}\BibitemShut {NoStop}%
\bibitem [{\citenamefont {Kermack}\ and\ \citenamefont
  {McKendrick}(1927)}]{Kermac1927}%
  \BibitemOpen
  \bibfield  {author} {\bibinfo {author} {\bibfnamefont {W.~O.}\ \bibnamefont
  {Kermack}}\ and\ \bibinfo {author} {\bibfnamefont {A.~G.}\ \bibnamefont
  {McKendrick}},\ }\href@noop {} {\bibfield  {journal} {\bibinfo  {journal}
  {Proc. R. Soc. Lond. A}\ }\textbf {\bibinfo {volume} {115}},\ \bibinfo
  {pages} {700} (\bibinfo {year} {1927})}\BibitemShut {NoStop}%
\bibitem [{\citenamefont {Chen}\ \emph {et~al.}(2013)\citenamefont {Chen},
  \citenamefont {Ghanbarnejad}, \citenamefont {Cai},\ and\ \citenamefont
  {Grassberger}}]{Chen2013}%
  \BibitemOpen
  \bibfield  {author} {\bibinfo {author} {\bibfnamefont {L.}~\bibnamefont
  {Chen}}, \bibinfo {author} {\bibfnamefont {F.}~\bibnamefont {Ghanbarnejad}},
  \bibinfo {author} {\bibfnamefont {W.}~\bibnamefont {Cai}}, \ and\ \bibinfo
  {author} {\bibfnamefont {P.}~\bibnamefont {Grassberger}},\ }\href {\doibase
  10.1209/0295-5075/104/50001} {\bibfield  {journal} {\bibinfo  {journal} {EPL
  (Europhysics Letters)}\ }\textbf {\bibinfo {volume} {104}},\ \bibinfo {pages}
  {50001} (\bibinfo {year} {2013})}\BibitemShut {NoStop}%
\bibitem [{\citenamefont {Janssen}\ and\ \citenamefont
  {Stenull}(2016)}]{Janssen2016}%
  \BibitemOpen
  \bibfield  {author} {\bibinfo {author} {\bibfnamefont {H.-K.}\ \bibnamefont
  {Janssen}}\ and\ \bibinfo {author} {\bibfnamefont {O.}~\bibnamefont
  {Stenull}},\ }\href {\doibase 10.1209/0295-5075/113/26005} {\bibfield
  {journal} {\bibinfo  {journal} {EPL (Europhysics Letters)}\ }\textbf
  {\bibinfo {volume} {113}},\ \bibinfo {pages} {26005} (\bibinfo {year}
  {2016})}\BibitemShut {NoStop}%
\bibitem [{\citenamefont {Cai}\ \emph {et~al.}(2015)\citenamefont {Cai},
  \citenamefont {Chen}, \citenamefont {Ghanbarnejad},\ and\ \citenamefont
  {Grassberger}}]{Chen2015}%
  \BibitemOpen
  \bibfield  {author} {\bibinfo {author} {\bibfnamefont {W.}~\bibnamefont
  {Cai}}, \bibinfo {author} {\bibfnamefont {L.}~\bibnamefont {Chen}}, \bibinfo
  {author} {\bibfnamefont {F.}~\bibnamefont {Ghanbarnejad}}, \ and\ \bibinfo
  {author} {\bibfnamefont {P.}~\bibnamefont {Grassberger}},\ }\href {\doibase
  10.1038/nphys3457} {\bibfield  {journal} {\bibinfo  {journal} {Nature
  physics}\ }\textbf {\bibinfo {volume} {11}},\ \bibinfo {pages} {936}
  (\bibinfo {year} {2015})}\BibitemShut {NoStop}%
\bibitem [{\citenamefont {Grassberger}\ \emph {et~al.}(2016)\citenamefont
  {Grassberger}, \citenamefont {Chen}, \citenamefont {Ghanbarnejad},\ and\
  \citenamefont {Cai}}]{Chen2016}%
  \BibitemOpen
  \bibfield  {author} {\bibinfo {author} {\bibfnamefont {P.}~\bibnamefont
  {Grassberger}}, \bibinfo {author} {\bibfnamefont {L.}~\bibnamefont {Chen}},
  \bibinfo {author} {\bibfnamefont {F.}~\bibnamefont {Ghanbarnejad}}, \ and\
  \bibinfo {author} {\bibfnamefont {W.}~\bibnamefont {Cai}},\ }\href {\doibase
  10.1103/PhysRevE.93.042316} {\bibfield  {journal} {\bibinfo  {journal} {Phys.
  Rev. E}\ }\textbf {\bibinfo {volume} {93}},\ \bibinfo {pages} {042316}
  (\bibinfo {year} {2016})}\BibitemShut {NoStop}%
\bibitem [{\citenamefont {Goltsev}\ \emph {et~al.}(2006)\citenamefont
  {Goltsev}, \citenamefont {Dorogovtsev},\ and\ \citenamefont
  {Mendes}}]{Goltsev2006}%
  \BibitemOpen
  \bibfield  {author} {\bibinfo {author} {\bibfnamefont {A.~V.}\ \bibnamefont
  {Goltsev}}, \bibinfo {author} {\bibfnamefont {S.~N.}\ \bibnamefont
  {Dorogovtsev}}, \ and\ \bibinfo {author} {\bibfnamefont {J.~F.~F.}\
  \bibnamefont {Mendes}},\ }\href {\doibase 10.1103/PhysRevE.73.056101}
  {\bibfield  {journal} {\bibinfo  {journal} {Phys. Rev. E}\ }\textbf {\bibinfo
  {volume} {73}},\ \bibinfo {pages} {056101} (\bibinfo {year}
  {2006})}\BibitemShut {NoStop}%
\bibitem [{\citenamefont {Parisi}\ and\ \citenamefont
  {Rizzo}(2008)}]{Parisi2008}%
  \BibitemOpen
  \bibfield  {author} {\bibinfo {author} {\bibfnamefont {G.}~\bibnamefont
  {Parisi}}\ and\ \bibinfo {author} {\bibfnamefont {T.}~\bibnamefont {Rizzo}},\
  }\href {\doibase 10.1103/PhysRevE.78.022101} {\bibfield  {journal} {\bibinfo
  {journal} {Phys. Rev. E}\ }\textbf {\bibinfo {volume} {78}},\ \bibinfo
  {pages} {022101} (\bibinfo {year} {2008})}\BibitemShut {NoStop}%
\bibitem [{\citenamefont {Sanz}\ \emph {et~al.}(2014)\citenamefont {Sanz},
  \citenamefont {Xia}, \citenamefont {Meloni},\ and\ \citenamefont
  {Moreno}}]{Sanz2014}%
  \BibitemOpen
  \bibfield  {author} {\bibinfo {author} {\bibfnamefont {J.}~\bibnamefont
  {Sanz}}, \bibinfo {author} {\bibfnamefont {C.-Y.}\ \bibnamefont {Xia}},
  \bibinfo {author} {\bibfnamefont {S.}~\bibnamefont {Meloni}}, \ and\ \bibinfo
  {author} {\bibfnamefont {Y.}~\bibnamefont {Moreno}},\ }\href {\doibase
  10.1103/PhysRevX.4.041005} {\bibfield  {journal} {\bibinfo  {journal} {Phys.
  Rev. X}\ }\textbf {\bibinfo {volume} {4}},\ \bibinfo {pages} {041005}
  (\bibinfo {year} {2014})}\BibitemShut {NoStop}%
\bibitem [{\citenamefont {Azimi-Tafreshi}(2016)}]{AzimiTafreshi2016}%
  \BibitemOpen
  \bibfield  {author} {\bibinfo {author} {\bibfnamefont {N.}~\bibnamefont
  {Azimi-Tafreshi}},\ }\href {\doibase 10.1103/PhysRevE.93.042303} {\bibfield
  {journal} {\bibinfo  {journal} {Phys. Rev. E}\ }\textbf {\bibinfo {volume}
  {93}},\ \bibinfo {pages} {042303} (\bibinfo {year} {2016})}\BibitemShut
  {NoStop}%
\bibitem [{\citenamefont {{Chen}}\ \emph {et~al.}(2016)\citenamefont {{Chen}},
  \citenamefont {{Ghanbarnejad}},\ and\ \citenamefont
  {{Brockmann}}}]{Brockmann2016}%
  \BibitemOpen
  \bibfield  {author} {\bibinfo {author} {\bibfnamefont {L.}~\bibnamefont
  {{Chen}}}, \bibinfo {author} {\bibfnamefont {F.}~\bibnamefont
  {{Ghanbarnejad}}}, \ and\ \bibinfo {author} {\bibfnamefont {D.}~\bibnamefont
  {{Brockmann}}},\ }\href@noop {} {\bibfield  {journal} {\bibinfo  {journal}
  {ArXiv e-prints}\ } (\bibinfo {year} {2016})},\ \Eprint
  {http://arxiv.org/abs/1603.09082} {arXiv:1603.09082 [physics.soc-ph]}
  \BibitemShut {NoStop}%
\bibitem [{\citenamefont {Catanzaro}\ \emph {et~al.}(2005)\citenamefont
  {Catanzaro}, \citenamefont {Bogu\~n\'a},\ and\ \citenamefont
  {Pastor-Satorras}}]{Catanzaro2005}%
  \BibitemOpen
  \bibfield  {author} {\bibinfo {author} {\bibfnamefont {M.}~\bibnamefont
  {Catanzaro}}, \bibinfo {author} {\bibfnamefont {M.}~\bibnamefont
  {Bogu\~n\'a}}, \ and\ \bibinfo {author} {\bibfnamefont {R.}~\bibnamefont
  {Pastor-Satorras}},\ }\href {\doibase 10.1103/PhysRevE.71.027103} {\bibfield
  {journal} {\bibinfo  {journal} {Phys. Rev. E}\ }\textbf {\bibinfo {volume}
  {71}},\ \bibinfo {pages} {027103} (\bibinfo {year} {2005})}\BibitemShut
  {NoStop}%
\bibitem [{\citenamefont {Newman}(2005{\natexlab{b}})}]{Newman2002}%
  \BibitemOpen
  \bibfield  {author} {\bibinfo {author} {\bibfnamefont {M.~E.~J.}\
  \bibnamefont {Newman}},\ }\enquote {\bibinfo {title} {Random graphs as models
  of networks},}\ in\ \href {\doibase 10.1002/3527602755.ch2} {\emph {\bibinfo
  {booktitle} {Handbook of Graphs and Networks}}}\ (\bibinfo  {publisher}
  {Wiley-VCH Verlag GmbH \& Co. KGaA},\ \bibinfo {year} {2005})\ pp.\ \bibinfo
  {pages} {35--68}\BibitemShut {NoStop}%
\bibitem [{\citenamefont {Pastor-Satorras}\ and\ \citenamefont
  {Vespignani}(2001)}]{PastorSatorras2001}%
  \BibitemOpen
  \bibfield  {author} {\bibinfo {author} {\bibfnamefont {R.}~\bibnamefont
  {Pastor-Satorras}}\ and\ \bibinfo {author} {\bibfnamefont {A.}~\bibnamefont
  {Vespignani}},\ }\href {\doibase 10.1103/PhysRevLett.86.3200} {\bibfield
  {journal} {\bibinfo  {journal} {Phys. Rev. Lett.}\ }\textbf {\bibinfo
  {volume} {86}},\ \bibinfo {pages} {3200} (\bibinfo {year}
  {2001})}\BibitemShut {NoStop}%
\bibitem [{\citenamefont {Moreno}\ \emph {et~al.}(2002)\citenamefont {Moreno},
  \citenamefont {Pastor-Satorras},\ and\ \citenamefont
  {Vespignani}}]{Moreno2002}%
  \BibitemOpen
  \bibfield  {author} {\bibinfo {author} {\bibfnamefont {Y.}~\bibnamefont
  {Moreno}}, \bibinfo {author} {\bibfnamefont {R.}~\bibnamefont
  {Pastor-Satorras}}, \ and\ \bibinfo {author} {\bibfnamefont {A.}~\bibnamefont
  {Vespignani}},\ }\href {\doibase 10.1140/epjb/e20020122} {\bibfield
  {journal} {\bibinfo  {journal} {The European Physical Journal B - Condensed
  Matter and Complex Systems}\ }\textbf {\bibinfo {volume} {26}},\ \bibinfo
  {pages} {521} (\bibinfo {year} {2002})}\BibitemShut {NoStop}%
\bibitem [{\citenamefont {Bogu{\~n}{\'a}}\ \emph {et~al.}(2003)\citenamefont
  {Bogu{\~n}{\'a}}, \citenamefont {Pastor-Satorras},\ and\ \citenamefont
  {Vespignani}}]{Boguna2003}%
  \BibitemOpen
  \bibfield  {author} {\bibinfo {author} {\bibfnamefont {M.}~\bibnamefont
  {Bogu{\~n}{\'a}}}, \bibinfo {author} {\bibfnamefont {R.}~\bibnamefont
  {Pastor-Satorras}}, \ and\ \bibinfo {author} {\bibfnamefont {A.}~\bibnamefont
  {Vespignani}},\ }in\ \href@noop {} {\emph {\bibinfo {booktitle} {Statistical
  Mechanics of Complex Networks}}},\ \bibinfo {series} {Lecture Notes in
  Physics}, Vol.\ \bibinfo {volume} {625},\ \bibinfo {editor} {edited by\
  \bibinfo {editor} {\bibfnamefont {R.}~\bibnamefont {Pastor-Satorras}},
  \bibinfo {editor} {\bibfnamefont {J.~M.}\ \bibnamefont {Rub{\'\i}}}, \ and\
  \bibinfo {editor} {\bibfnamefont {A.~D.-G.}\ \bibnamefont {a}}}\ (\bibinfo
  {publisher} {Springer Verlag},\ \bibinfo {address} {Berlin},\ \bibinfo {year}
  {2003})\ pp.\ \bibinfo {pages} {127--147}\BibitemShut {NoStop}%
\bibitem [{\citenamefont {Callaway}\ \emph {et~al.}(2000)\citenamefont
  {Callaway}, \citenamefont {Newman}, \citenamefont {Strogatz},\ and\
  \citenamefont {Watts}}]{Callaway2000}%
  \BibitemOpen
  \bibfield  {author} {\bibinfo {author} {\bibfnamefont {D.~S.}\ \bibnamefont
  {Callaway}}, \bibinfo {author} {\bibfnamefont {M.~E.~J.}\ \bibnamefont
  {Newman}}, \bibinfo {author} {\bibfnamefont {S.~H.}\ \bibnamefont
  {Strogatz}}, \ and\ \bibinfo {author} {\bibfnamefont {D.~J.}\ \bibnamefont
  {Watts}},\ }\href {\doibase 10.1103/PhysRevLett.85.5468} {\bibfield
  {journal} {\bibinfo  {journal} {Phys. Rev. Lett.}\ }\textbf {\bibinfo
  {volume} {85}},\ \bibinfo {pages} {5468} (\bibinfo {year}
  {2000})}\BibitemShut {NoStop}%
\bibitem [{\citenamefont {Dorogovtsev}\ \emph {et~al.}(2008)\citenamefont
  {Dorogovtsev}, \citenamefont {Goltsev},\ and\ \citenamefont
  {Mendes}}]{Dorogovtsev2008}%
  \BibitemOpen
  \bibfield  {author} {\bibinfo {author} {\bibfnamefont {S.~N.}\ \bibnamefont
  {Dorogovtsev}}, \bibinfo {author} {\bibfnamefont {A.~V.}\ \bibnamefont
  {Goltsev}}, \ and\ \bibinfo {author} {\bibfnamefont {J.~F.~F.}\ \bibnamefont
  {Mendes}},\ }\href {\doibase 10.1103/RevModPhys.80.1275} {\bibfield
  {journal} {\bibinfo  {journal} {Rev. Mod. Phys.}\ }\textbf {\bibinfo {volume}
  {80}},\ \bibinfo {pages} {1275} (\bibinfo {year} {2008})}\BibitemShut
  {NoStop}%
\bibitem [{\citenamefont {Newman}(2002)}]{Newman2002b}%
  \BibitemOpen
  \bibfield  {author} {\bibinfo {author} {\bibfnamefont {M.~E.~J.}\
  \bibnamefont {Newman}},\ }\href {\doibase 10.1103/PhysRevE.66.016128}
  {\bibfield  {journal} {\bibinfo  {journal} {Phys. Rev. E}\ }\textbf {\bibinfo
  {volume} {66}},\ \bibinfo {pages} {016128} (\bibinfo {year}
  {2002})}\BibitemShut {NoStop}%
\end{thebibliography}%

\end{document}